\documentclass[12pt]{article}
\usepackage[a4paper, total={7in, 10in}]{geometry}
\usepackage[parfill]{parskip}
\usepackage{physics, tensor, float, subcaption}
\usepackage{graphicx}
\graphicspath{ {Plots/} }
\usepackage{jhep-mod} 
\usepackage{bm}
\usepackage{soul}
\usepackage{amssymb,amsmath,amsthm}
\usepackage{mathrsfs}
\usepackage[utf8]{inputenc}
\usepackage{enumerate}
\usepackage{bigints}
\usepackage{xcolor}
\usepackage{appendix}
\usepackage{graphicx}
\usepackage{float}
\usepackage{tikz}
\usepackage{setspace}
\usepackage{cancel}
\usepackage{tocbasic}
\DeclareTOCStyleEntry[
  beforeskip=.2em plus 1pt,
  pagenumberformat=\textbf
]{tocline}{section}
\definecolor{purple}{rgb}{1,0,1}
\definecolor{lime}{HTML}{A6CE39} 

\definecolor{lime}{HTML}{A6CE39}
\newcommand{\orcidicon}{%
	\begin{tikzpicture}
	\draw[lime, fill=lime] (0,0) 
		circle [radius=0.16] 
		node[white] {{\fontfamily{qag}\selectfont \tiny ID}};
	\draw[white, fill=white] (-0.0625,0.095) 
		circle [radius=0.007];
	\end{tikzpicture}
	\hspace{-5mm}
}
\newcommand\orcidJoshua{{\href{https://orcid.org/0000-0003-1200-7261}{\orcidicon}}}
\newcommand\orcidThomas{{\href{https://orcid.org/0000-0002-0314-4136}{\orcidicon}}}
\newcommand\orcidAlex{{\href{https://orcid.org/0000-0002-1763-3563}{\orcidicon}}}
\newcommand\orcidMatt{{\href{https://orcid.org/0000-0003-1088-6485}{\orcidicon}}}

\begin{document}

\title{
\huge{
\leftline{Unit-lapse versions of the Kerr spacetime}
}}
\author{
\Large
Joshua Baines\!\orcidJoshua\!, Thomas Berry\!\orcidThomas\!, Alex Simpson\!\orcidAlex\!,
\\{\sf  and} Matt Visser\!\orcidMatt}
\affiliation{School of Mathematics and Statistics, Victoria University of Wellington, 
\\
\null\qquad PO Box 600, Wellington 6140, New Zealand.}
\emailAdd{joshua.baines@sms.vuw.ac.nz}
\emailAdd{thomas.berry@sms.vuw.ac.nz}
\emailAdd{alex.simpson@sms.vuw.ac.nz}
\emailAdd{matt.visser@sms.vuw.ac.nz}

\abstract{
\vspace{1em}

The Kerr spacetime is perhaps the most astrophysically important of the currently known exact solutions to the Einstein field equations. 
Whenever spacetimes can be put in unit-lapse form it becomes possible to identify some very  straightforward timelike geodesics, (the ``rain'' geodesics), making the physical interpretation of these spacetimes particularly clean and elegant. The most well-known of these unit-lapse formulations is the Painlev\'e--Gullstrand form of the Schwarzschild spacetime, though there is also a Painlev\'e--Gullstrand form of the Lense--Thirring (slow rotation) spacetime. More radically there are also two known unit-lapse forms of the Kerr spacetime --- the Doran and Nat\'ario metrics --- though these are not precisely in Painlev\'e--Gullstrand form. Herein we shall seek to explicate the most general unit-lapse form of the Kerr spacetime. 
While at one level this is ``merely'' a choice of coordinates, it is a strategically and tactically useful choice of coordinates, thereby making the technically challenging but astrophysically crucial Kerr spacetime somewhat easier to deal with. 

\bigskip
\noindent
{\sc Date:}  10 August 2020; \LaTeX-ed \today


\bigskip
\noindent{\sc PhySH:} 
Gravitation
}

\maketitle
\def\tr{{\mathrm{tr}}}
\def\diag{{\mathrm{diag}}}
\def\cof{{\mathrm{cof}}}
\def\pdet{{\mathrm{pdet}}}
\parindent0pt
\parskip7pt

\vspace{-25pt}
\section{Introduction}
The Kerr spacetime~\cite{Kerr,Kerr-Texas,kerr-intro,kerr-book,kerr-book-2,Adler-Bazin-Schiffer,D'Inverno,Hartle,Carroll,wald,weinberg, Hobson, MTW}
is perhaps the most astrophysically important of the known exact solutions to the Einstein field equations. 
Many physically interesting spacetimes, (both theoretically interesting and astrophysically interesting),  can be put in unit-lapse form. 
That is, for many physically  interesting spacetimes  one can find coordinate charts such that the ADM foliation~\cite{MTW},
which generally entails a metric decomposition of the form 
\begin{equation}
g_{ab} = \left[ \begin{array}{c|c} -N^2 + (h^{ij} v_i v_j) & -v_j \\  \hline -v_i & h_{ij} \end{array}\right]_{ab};
\qquad \qquad
g^{ab} = \left[ \begin{array}{c|c} -N^{-2}    & -v^j  N^{-2} \\  \hline -v^i N^{-2} & h^{ij} - v^i v^j N^{-2}\end{array}\right]^{ab};
\end{equation}
can instead be specialized to
\begin{equation}
g_{ab} = \left[ \begin{array}{c|c} -1 + (h^{ij} v_i v_j) & -v_j \\  \hline -v_i & h_{ij} \end{array}\right]_{ab};
\qquad \qquad
g^{ab} = \left[ \begin{array}{c|c} -1   & -v^j \\  \hline -v^i & h^{ij} - v^i v^j\end{array}\right]^{ab}.
\end{equation}
Here $h^{ij} = [h_{ij}]^{-1}$ and $v^i = h^{ij} \, v_j$. 
Our signature is $-+++$. Space-time indices such as $a$, $b$, $c$, $d$ run 0\dots 3, with $x^0=t$, while spatial
indices such as $i$, $j$, $k$, $l$ run 1\dots 3. 
Physically $h_{ij}$ is interpreted as the 3-metric of the constant-$t$ spatial slices, while the flow vector $v_i$ is the negative of what is usually called the shift vector. 
The unit-lapse condition $N\to1$ is encoded in the statement that $g^{tt}=-1$, or equivalently that $\det(g_{ab})=-\det(h_{ij})$. 
Equivalently one can write the unit-lapse line-element as:
\begin{equation}
ds^2 = - dt^2 + h_{ij} (dx^i - v^i dt) (dx^j-v^j dt).
\end{equation}
Once one has the metric presented in unit-lapse form, the ``rain'' geodesics (timelike geodesics corresponding to test particles dropped from spatial infinity with zero initial velocity) are particularly simple and give clean mathematically and physically transparent insight into the spacetime geometry~\cite{river}.

Spacetimes that can be put in this unit-lapse form include the Painlev\'e--Gullstrand form of the Schwarzschild spacetime~\cite{schwarzschild-1916, painleve1, painleve2, gullstrand,poisson}
\begin{eqnarray}
\label{E:PG-Sch}
d s^2 &=& - d t^2 +\left(d r+\sqrt{2m\over r} \; dt\right)^2
+ r^2 \left(d\theta^2+\sin^2\theta\; d\phi^2\right),
\end{eqnarray}
the Painlev\'e--Gullstrand form of the Lense--Thirring
spacetime~\cite{Lense-Thirring, Pfister, 4-of-us}
\begin{eqnarray}
\label{E:LT5}
d s^2 &=& - d t^2 +\left(d r+\sqrt{2m\over r} \; dt\right)^2
+ r^2 \left(d\theta^2+\sin^2\theta\; \left(d\phi - {2J\over r^3} dt\right) ^2\right),
\end{eqnarray}
and, [at least for $r\geq Q^2/(2m)$], the Painlev\'e--Gullstrand form of the Reissner--Nordstr\"om spacetime
\begin{eqnarray}
\label{E:PG-RN2}
d s^2 &=& - d t^2 +\left(d r+\sqrt{{2m\over r}-{Q^2\over r^2}} \; dt\right)^2
+ r^2 \left(d\theta^2+\sin^2\theta\; d\phi^2\right).
\end{eqnarray}
More subtly there are already at least two known distinct unit-lapse forms of the Kerr spacetime, the fully explicit Doran metric~\cite{doran}, and the semi-explicit Nat\'ario metric~\cite{natario}. 
(For considerable general background on the Kerr spacetime geometry see the technical references~\cite{Kerr, Kerr-Texas, kerr-intro, kerr-book, kerr-book-2}, 
and the textbooks~\cite{Adler-Bazin-Schiffer, D'Inverno, Hartle, Carroll, wald, weinberg, Hobson, MTW}.) 
\enlargethispage{20pt}

Herein we shall develop several additional and particularly simple unit-lapse variants of the Kerr spacetime. We shall compare and contrast them with the  Doran~\cite{doran} and Nat\'ario~\cite{natario} metrics, and generalize them by embedding them in what we shall argue is the most general unit-lapse representation of the Kerr spacetime. 
While at one level this is ``merely'' a choice of coordinates, it is a strategically and tactically useful choice of coordinates, making the technically challenging but astrophysically crucial Kerr spacetime somewhat easier to deal with. 

It is also worth noting that unit-lapse spacetimes occur quite commonly and naturally in many examples of analogue spacetimes~\cite{unexpected, visser:1997,visser:1998,volovik:1999,stone:2001,visser:2001,fischer:2002, novello:2002,probing,Visser:vortex,LRR,Liberati:2005,Weinfurtner:2005, visser:2010,Visser:2013,Liberati:2018,Schuster:2018}  --- where the unit lapse condition physically corresponds to a constant propagation speed, (for example, sound waves in water). 
So various analogue spacetimes can be invoked to develop physical intuition in this purely general relativistic context. 

\section{``Rain'' geodesics}

Whenever one has a metric presented in unit-lapse form, at least some of the timelike geodesics, the ``rain'' geodesics corresponding to a test object being dropped from spatial infinity with zero initial velocity, are particularly easy to analyze~\cite{river}. 
Consider the contravariant vector field
\begin{equation}
V^a = - g^{ab} \,\nabla_b t = - g^{ta} = \left(1; v^i\right).
\end{equation}
The corresponding covariant vector field is
\begin{equation}
V_a = -  \nabla_a t =  \left(-1;0, 0,0\right).
\end{equation}
Thence $g_{ab}\, V^a V^b = V^a V_a = -1$, so $V^a$ is a  future-pointing timelike vector field with unit norm, a 4-velocity. 
But this vector field has zero 4-acceleration:
\begin{equation}
A_a = V^b \nabla_b V_a = - V^b \nabla_b \nabla_a t = - V^b \nabla_a \nabla_b t 
=  V^b \nabla_a V_b = {1\over2} \nabla_a (V^b V_b) = 0.
\end{equation}
Thus the integral curves of $V^a$ are timelike \emph{geodesics}. 
Specifically, the  integral curves represented by
\begin{equation}
{dx^a\over d\tau} = \left( {dt\over d\tau};{dx^i\over d\tau}\right) =
\left( 1 ;  v^i\right) 
\end{equation}
are timelike geodesics.
Integrating the first of these equations is trivial
\begin{equation}
t(\tau) = \tau;
\end{equation}
so that the time coordinate $t$ can be identified with the proper time of these particular geodesics.
The remaining three equations,
\begin{equation}
{dx^i\over dt} = v^i(x),
\end{equation}
will depend on the specific form of the flow vector $v^i(x)$, and we will explore them more carefully (perhaps exhaustively) in the analysis below.

\section{Coordinate transformations}

The Kerr spacetime is both stationary and axisymmetric~\cite{Kerr,Kerr-Texas,kerr-intro,kerr-book,kerr-book-2,Adler-Bazin-Schiffer, D'Inverno,Hartle,Carroll,wald,weinberg, Hobson, MTW}. 
Let us label the coordinates as $(t,r,\theta,\phi)$.
Using the symmetries of the Kerr spacetime  it is possible to set up preferred temporal and axial coordinates $t$ and $\phi$ to make the relevant Killing vectors simple:
\begin{equation}
K^a=(1,0,0,0)^a; \qquad \hbox{and} \qquad \tilde K^a = (0,0,0,1)^a.
\end{equation}
As is completely standard, the metric components then satisfy $\partial_t g_{ab}=0=\partial_\phi g_{ab}$.

\subsection{Symmetry-preserving coordinate transformations}
If one now restricts one's attention to coordinate transformations that do not disturb these nice features of the presentation,
 (that is, coordinate transformations that keep the stationary and axisymmetric symmetries \emph{manifest}), 
 one is forced to specialize to coordinate transformations of the form
\begin{equation}
t \to \bar t = t + T(r,\theta); \qquad \phi \to \bar\phi = \phi + \Phi(r,\theta); \qquad
\end{equation}
\begin{equation}
(r,\theta) \to (\bar r, \bar\theta) = \left(\bar r(r,\theta), \bar \theta(r,\theta)\right). 
\end{equation}
For current purposes we shall leave the $r$ and $\theta$ coordinates intact, 
and shall further specialise to coordinate transformations affecting $t$ and $\phi$ only. 
One then has
\begin{equation}
dt \to d\bar t = dt + T_r\,dr +T_\theta\,d\theta ; \qquad 
d\phi \to d\bar\phi = d\phi + \Phi_r\,dr +\Phi_\theta\,d\theta.
\end{equation}
The relevant Jacobi matrix is
\begin{equation}
J^a{}_b = {\partial \bar x^a\over \partial x^b} =
\left.\left[ \begin{array}{cccc}
1 & T_r & T_\theta &0 \\
0&1&0&0\\
0&0&1&0\\
0& \Phi_r &\Phi_\theta & 1
\end{array}\right]^a\right._b;
\qquad
\det(J^a{}_b)=1.
\end{equation}

\subsection{Temporal-only coordinate transformations}
Let us first consider $t$-only coordinate transformations, leaving $\phi$ fixed. 
The Jacobi matrix reduces to
\begin{equation}
J^a{}_b = {\partial \bar x^a\over \partial x^b} =
\left.\left[ \begin{array}{cccc}
1 & T_r & T_\theta &0 \\
0&1&0&0\\
0&0&1&0\\
0& 0&0 & 1
\end{array}\right]^a\right._b.
\end{equation}
For the inverse metric we then have
\begin{equation}
\bar g^{ab} = J^a{}_c\; J^b{}_d\; g^{cd}.
\end{equation}
Specifically
\begin{equation}
\bar g^{tt} = J^t{}_c\, J^t{}_d\, g^{cd} = g^{tt} + 2 T_i g^{ti} + T_i T_j g^{ij} = 
- N^{-2} (1+ v^i T_i)^2 + h^{ij} T_i T_j.
\end{equation}
So to enforce unit lapse, $\bar g^{tt} \to -1$, \emph{if it can be done at all}, one needs to solve the partial differential equation (PDE):
\begin{equation}
-1  = g^{tt} + 2 T_i \; g^{ti} + T_i T_j \; g^{ij} .
\end{equation}
Equivalently, one needs to solve
\begin{equation}
-1 =  - N^{-2} (1+ v^i \; T_i)^2 + h^{ij} \;T_i T_j,
\end{equation}
to find the function $T(r,\theta)$ specifying the  transformation of the $t$ coordinate. 

Whether or not this PDE can be solved depends on specific features of the underlying spacetime. 
For instance, spherical symmetry will certainly do the job, since then $T(r)$ is a function of $r$ only, and we simply need to solve a quadratic equation for $T_r$:
\begin{equation}
-1 =  - N^{-2} (1+ v^r\;T_r)^2 + h^{rr} \; T_r^2.
\end{equation}
Furthermore, as we shall soon see, in the specific situation we are interested in, special features of the Kerr spacetime will do the job as well.

Note that simplifying the lapse generally makes other parts of the metric tensor more complicated.
Consider the flow vector; we note that in general
\begin{equation}
\bar v^i = - \bar g^{ti} = - J^t{}_c\, J^i{}_d\, g^{cd} 
= - J^t{}_t\, J^i{}_t\, g^{tt}  - J^t{}_t\, J^i{}_j\, g^{tj} -  J^t{}_k\, J^i{}_t\, g^{kt} - J^t{}_k\, J^i{}_l\, g^{kl}.
\end{equation}
But since in the present situation $J^t{}_t =1$, $J^i{}_t=0$, $J^t{}_i = T_i$, and $J^i{}_j = \delta^i{}_j$, this collapses to
\begin{equation}
\bar v^i
=  - \, g^{ti}- J^t{}_j \,g^{ij} = {v^i\over N^2} - T_j\left (h^{ij} - {v^i v^j\over N^2} \right)
= v^i \left(1+T_j \,v^j\over N^2\right) - h^{ij} \,T_j.
\end{equation}
That is, the coordinate transformation that simplifies the lapse to unity will also modify (and typically complicate) the flow vector. 

Furthermore, for the 3-metric
\begin{equation}
\bar h_{ij} = \bar g_{ij} = J^a{}_i\;J^b{}_j\; g_{ab} = g_{ij} + g_{it} T_j + g_{jt} T_i + g_{tt} T_i T_j.
\end{equation}
This implies
\begin{equation}
\bar h_{ij}
= h_{ij} - v_i T_j - T_i v_j - (N^2-(h_{kl} v^k v^l)) \; T_i T_j. 
\end{equation}
That is, the coordinate transformation that simplifies the lapse to unity will also modify (and typically complicate) the 3-metric. 

\subsection{Azimuthal-only coordinate transformations}

Now assume for the sake of argument that one has successfully used the freedom to choose the function $T(r,\theta)$ to put the metric into unit lapse form, $N\to1$. 
What more can be done by now using the $\phi$ transformation and the function $\Phi(r,\theta)$? 
We are now interested in keeping the $t$ coordinate fixed and considering
\begin{equation}
J^a{}_b = {\partial \bar x^a\over \partial x^b} =
\left.\left[ \begin{array}{cccc}
1 & 0 & 0 &0 \\
0&1&0&0\\
0&0&1&0\\
0& \Phi_r &\Phi_\theta & 1
\end{array}\right]^a\right._b.
\end{equation}
We note that this coordinate transformation will not disturb the unit-lapse condition, whereas for the flow vector
\begin{equation}
\bar v^i = - \bar g^{ti} = - J^t{}_c\, J^i{}_d\, g^{cd} 
= - J^t{}_t\, J^i{}_t\, g^{tt}  - J^t{}_t\, J^i{}_j\, g^{tj} -  J^t{}_k\, J^i{}_t\, g^{kt} - J^t{}_k\, J^i{}_l\, g^{kl}.
\end{equation}
But since in the current situation $J^i{}_t=0= J^t{}_i$ this collapses to
\begin{equation}
\bar v^i
=  - J^i{}_j\, g^{tj} =  J^i{}_j\, v^j  = v^i + \left(0,\;0,\;\Phi_r v^r + \Phi_\theta v^\theta\right)^i.
\end{equation}
That is, $\bar v^r=v^r$, $\bar v^\theta=v^\theta$, but $\bar v^\phi = v^\phi +\Phi_r v^r + \Phi_\theta v^\theta$.  
So we can use the remaining coordinate freedom in $\phi$ to attempt to simplify the contravariant $\phi$ component of the flow vector. 
Doing so would then simplify the rain geodesics. 
Of course there is a price to pay: For the inverse 3-metric one now has
\begin{equation}
\bar g^{ij} = J^i{}_a\;J^j{}_b\; g^{ab} = J^i{}_k\;J^j{}_l\; g^{kl},
\label{E:inverse-metric-spatial}
\end{equation}
implying
\begin{equation}
\bar g^{rr}=g^{rr}; \qquad \bar g^{r\theta}=g^{r\theta}; \qquad
\bar g^{\theta\theta} = g^{\theta\theta};
\label{E:inverse-metric-spatial-1}
\end{equation}
\begin{equation}
\bar g^{r\phi}=g^{r\phi}+ g^{rr} \Phi_r + g^{r\theta} \Phi_\theta; \qquad 
\bar g^{\theta\phi}=g^{\theta\phi}+g^{\theta r} \Phi_r + g^{\theta\theta} \Phi_\theta; 
\label{E:inverse-metric-spatial-2}
\end{equation}
\begin{equation}
\bar g^{\phi\phi}=g^{\phi\phi}+ (g^{\phi r}\Phi_r + g^{\phi\theta} \Phi_\theta) 
+ (g^{rr} \Phi_r^2 + 2 g^{r\theta} \Phi_r \Phi_\theta + g^{\theta\theta} \Phi_\theta^2 ).
\label{E:inverse-metric-spatial-3}
\end{equation}
That is, the coordinate transformation that (potentially) simplifies the flow vector will also modify (and typically complicate) the 3-metric. 
\enlargethispage{20pt}
The arguments presented so far have been rather general, appealing merely to stationarity and axisymmetry.
Let us now see how these considerations apply in the specific case of the Kerr spacetime.

\section{Enforcing unit lapse --- rain metrics for Kerr spacetime}

Let us first focus on two particularly simple and novel unit-lapse versions of the Kerr spacetime, based on Boyer--Lindquist and Eddington--Finkelstein coordinates respectively. 

\subsection{Boyer--Lindquist-rain metric}
\def\BLrain{{\hbox{\tiny BL-rain}}}

The Kerr line element in the usual Boyer--Lindquist coordinates is
\begin{eqnarray}
(ds^2)_{BL}&=& -\left(1-{2mr\over\rho^2}\right)dt^2 -{4mar\sin^2\theta\over\rho^2} d\phi dt + {\rho^2\over\Delta} dr^2 + \rho^2d\theta^2
\nonumber\\
&& +\left( r^2+a^2+ {2mra^2\sin^2\theta\over\rho^2}\right) \sin^2\theta d\phi^2,
\end{eqnarray}
with the usual definitions $\rho=\sqrt{r^2+a^2\cos^2\theta}$ and $\Delta= r^2+a^2-2mr$.
Some authors instead use the notation $\Sigma = r^2+a^2\cos^2\theta$, which we find to be not useful and shall avoid.
Other authors prefer to define
\begin{equation}
\Sigma = r^2+a ^2 +{2mra^2\over\rho^2} \sin^2\theta
=\rho^2 + a^2 \left(1+{2mr\over\rho^2} \right)\sin^2\theta,
\end{equation}
which we find to be more useful.
\enlargethispage{40pt}

Thence for the covariant Boyer--Lindquist metric\footnote{It is useful to note that as $a\to0$ one regains Schwarzschild spacetime in the usual curvature coordinates.} 
\begin{equation}
(g_{ab})_{BL} =  {\left[\begin{array}{c|cc|c}
- 1+{2mr\over\rho^2} & 0 &
 0 & -{2mar\sin^2\theta\over \rho^2}\\ \hline
 0&{\rho^2\over \Delta}
 &0&0\\
 0&0&{\rho^2}&0\\ \hline
- {2mar\sin^2\theta\over \rho^2}&0&
0& 
\Sigma \;\sin^2\theta
\end{array}\right]}_{ab},
\end{equation}
and
\begin{equation}
\det\left[ (g_{ab})_{BL} \right] = -\rho^4 \sin^2\theta.
\end{equation}
Furthermore it is an easy exercise to check that the inverse metric is \begin{equation}
(g^{ab})_{BL} =  {\left[\begin{array}{c|cc|c}
- 1-{2mr(r^2+a^2)\over\rho^2\Delta} & 0 &
 0 & -{2mar\over \rho^2\Delta}\\ \hline
 0&{\Delta\over \rho^2}
 &0&0\\
 0&0&{1\over \rho^2}&0\\ \hline
- {2mar\over \rho^2\Delta}&0&
0& 
{1-2mr/\rho^2\over \Delta\sin^2\theta}
\end{array}\right]}^{ab}.
\end{equation}
In fact we shall soon see that in the Kerr spacetime the inverse (contravariant) metric is often simpler than the (covariant) metric itself.

\clearpage
Working slowly and carefully for clarity, we recall that to put this into unit lapse form we would need to solve
\begin{equation}
-1 =  - N^{-2} (1+ v^i T_i)^2 + h^{ij} T_i T_j.
\end{equation}
Noting that in this current situation $v^i \,T_i=0$, this equation reduces to
\begin{equation}
N^{-2} -1 =  h^{ij} T_i T_j.
\end{equation}
That is
\begin{equation}
{2mr(r^2+a^2)\over\rho^2\Delta} =  
\left({ \Delta\over \rho^2} \;T_r^2 + {1\over \rho^2} \;T_\theta^2\right).
\end{equation}
Thence, multiplying through by $\rho^2$ we see
\begin{equation}
{2mr(r^2+a^2)\over\Delta} =  
\left({ \Delta}\; T_r^2 + T_\theta^2\right).
\end{equation}
But this has the obvious solutions
\begin{equation}
T_\theta=0; \qquad  T_r = \pm {\sqrt{2mr(r^2+a^2)}\over\Delta}.
\end{equation}
So $T(r,\theta)$ is actually independent of $\theta$, and we explicitly have
\begin{equation}
T(r) = \pm \int {\sqrt{2mr(r^2+a^2)}\over\Delta} dr.
\end{equation}
Thence
\begin{equation}
\bar t = t + T(r); \qquad d\bar t = dt + T_r; \qquad dt = d\bar t - T_r.
\end{equation}
That is, now suppressing the overbar,
simply taking the Boyer--Lindquist form of the Kerr metric and replacing
\begin{equation}
dt \to dt \mp  {\sqrt{2mr(r^2+a^2)}\over\Delta}\; dr,
\end{equation}
will put the metric into unit-lapse form. 
There are two roots, and retrospectively checking that one has a black hole (rather than a white hole) leads one to choose the {negative} root.\footnote{This is most easily checked by setting $a\to0$ and comparing with the (black hole) Painlev\'e--Gullstrand form of the Schwarschild line element.}
Let us call the resulting line element the Boyer--Lindquist-rain metric, also to be abbreviated as the BL-rain metric.

We have
\begin{eqnarray}
(ds^2)_\BLrain&=& 
-\left(1-{2mr\over\rho^2}\right)\left(dt {-}  {\sqrt{2mr(r^2+a^2)}\over\Delta} \; dr\right)^2 
\nonumber\\
&& 
-{4mar\sin^2\theta\over\rho^2} d\phi 
\left(dt {-}  {\sqrt{2mr(r^2+a^2)}\over\Delta}\; dr\right) + {\rho^2\over\Delta} dr^2 + \rho^2d\theta^2
\nonumber\\
&& 
+\Sigma \sin^2\theta d\phi^2.
\end{eqnarray}
Thence 
we have the somewhat messy result that the covariant metric $(g_{ab})_\BLrain$ equals
\begin{equation}
{\left[\begin{array}{c|cc|c}
- 1+{2mr\over\rho^2} & \left(1-{2mr\over\rho^2}\right) {\sqrt{2mr(r^2+a^2)}\over\Delta} &
 0 & -{2mar\sin^2\theta\over \rho^2}\\ \hline
 \left(1-{2mr\over\rho^2}\right) {\sqrt{2mr(r^2+a^2)}\over\Delta}&
{\rho^2\over \Delta} - \left(1-{2mr\over\rho^2}\right) {2mr(r^2+a^2)\over\Delta^2}
 &0&+{2mar\sin^2\theta\over \rho^2}  {\sqrt{2mr(r^2+a^2)}\over\Delta}  \\
 0&0&{\rho^2}&0\\ \hline
- {2mar\sin^2\theta\over \rho^2}&+{2mar\sin^2\theta\over \rho^2}  {\sqrt{2mr(r^2+a^2)}\over\Delta} &
0& 
\Sigma\sin^2\theta
\end{array}\right]}_{ab}
\end{equation}
while we still retain the simple result that
\begin{equation}
\det\left[ (g_{ab})_\BLrain \right] = -\rho^4 \sin^2\theta.
\end{equation}
Furthermore, it is an easy exercise to check that the inverse metric now takes on the relatively simple form
\begin{equation}
(g^{ab})_\BLrain =  {\left[\begin{array}{c|cc|c}
- 1 & {\sqrt{2mr(r^2+a^2)}\over \rho^2}  &
 0 & -{2mar\over \rho^2\Delta}\\ \hline
 {\sqrt{2mr(r^2+a^2)}\over \rho^2} &{\Delta\over \rho^2}
 &0&0\\
 0&0&{1\over \rho^2}&0\\ \hline
- {2mar\over \rho^2\Delta}&0&
0& 
{1-2mr/\rho^2\over\Delta\sin^2\theta}
\end{array}\right]}^{ab}.
\end{equation}
So we have indeed simplified the lapse, but at the cost of complicating the flow vector:
\begin{equation}
N=1; \qquad (v^i)_\BLrain = \left(-{\sqrt{2mr(r^2+a^2)}\over \rho^2} ,\;
 0 ,\; {2mar\over \rho^2\Delta}\right).
\end{equation}
Note that for the rain geodesics $d\theta/dt=0$, so that $\theta(t)=\theta_\infty$ is conserved. Also
\begin{equation}
\left({d\phi\over dr}\right)_\BLrain = {d\phi/dt\over dr/dt} =
-{ a\sqrt{2mr}\over\Delta\sqrt{r^2+a^2}}. 
\end{equation}
Therefore for these BL-rain geodesics we have
\begin{equation}
\phi(r) = \phi_\infty + \int_r^\infty { a\sqrt{2mr}\over\Delta\sqrt{r^2+a^2}}\; dr.
\end{equation}
Overall this BL-rain version of the Kerr spacetime is quite straightforward, both  in terms of tractability and clarity of physical insight.

\subsection{Eddington--Finkelstein-rain metric}
\def\EFnull{{\hbox{\tiny EF-null}}}
\def\EFtr{{\hbox{\tiny EF-tr}}}
\def\EFrain{{\hbox{\tiny EF-rain}}}
\def\d{d} 

The very first version of the Kerr spacetime, as presented in Kerr's original PRL article~\cite{Kerr}, was in terms of 
Eddington--Finkelstein null coordinates (note the \emph{sign} of the parameter $a$ has been flipped in order to conform to standard conventions). We shall abbreviate the name of this metric as EF-null: 
\begin{eqnarray}
(d s^2)_\EFnull &=& -\left[ 1 - {2mr\over \rho^2}\right]\; \left(d u - a \sin^2\theta \; d \phi\right)^2
\nonumber
\\
&&
+2 \left(d u - a \sin^2\theta \; d \phi\right) \; \left(d r - a \sin^2\theta \; d \phi\right)
\nonumber
\\
&&
+ \rho^2\; (d\theta^2+\sin^2\theta\;d\phi^2).
\label{E:K1}
\end{eqnarray}
First, consider a slightly different but completely equivalent form of the metric which follows from Kerr's original ``advanced Eddington--Finkelstein'' form via the coordinate substitution
\begin{equation}
\label{E:BL-change-1}
u = t+r, \qquad du = dt + dr,
\end{equation}
in which case we have what we shall abbreviate as the EF-tr line element:
\begin{eqnarray}
\label{E:K3}
(\d s^2)_\EFtr &=& 
-\d t^2
+ \d r^2 - 2 a \sin^2\theta \;\d r \; \d \phi + \rho^2\; \d\theta^2 
+ (r^2+a^2) \sin^2\theta \;\d\phi^2
\nonumber
\\
&&
+{2mr\over \rho^2}\; \left(\d t + \d r - a \sin^2\theta \; \d \phi\right)^2.
\end{eqnarray}
Note that with this sign convention for the parameter $a$ one has the standard Lense--Thirring result for weak fields at large distances~\cite{Lense-Thirring,Pfister,4-of-us}. 
Also note that if $a\to0$ then this reduces to the Eddington--Finkelstein $t$-$r$ form of Schwarzschild spacetime.
Keeping $a\neq 0$, in these Eddington--Finkelstein $t$-$r$ coordinates
the covariant metric $(g_{ab})_\EFtr$ is 
\begin{equation}
(g_{ab})_\EFtr = 
 {\left[\begin{array}{c|cc|c}
- 1+{2mr\over \rho^2} & {2mr\over \rho^2} &
 0 &- {2mar\over\rho^2}\sin^2\theta \\ \hline
 {2mr\over\rho^2}&1+{2mr\over\rho^2}
 &0&-a(1+{2mr\over\rho^2})\sin^2\theta\\ 
 0&0&\;\;\rho^2\;\;&0\\ \hline
 -{2mar\over\rho^2}\sin^2\theta&-a(1+{2mr\over\rho^2})\sin^2\theta&0& 
 \Sigma\, \sin^2\theta
\end{array}\right]}_{ab}.
\end{equation}

\clearpage
In contrast, in these Eddington--Finkelstein $t$-$r$ coordinates the Kerr geometry has the rather simple inverse metric
\begin{equation}
(g^{ab})_\EFtr =  {\left[\begin{array}{c|cc|c}
- 1-{2mr\over \rho^2} & {2mr\over \rho^2} &
 0 & 0\\ \hline
 {2mr\over \rho^2}&{\Delta\over \rho^2}
 &0& {a\over \rho^2}\\
 0&0&{1\over \rho^2}&0\\ \hline
 0& {a\over \rho^2}&0& {1\over \rho^2\sin^2\theta}
\end{array}\right]}^{ab}.
\end{equation}
To put this into unit lapse form we would need to solve the PDE
\begin{equation}
-1  = g^{tt} + 2 T_i \;g^{ti} + T_i T_j \;g^{ij}.
\end{equation}
That is
\begin{equation}
-1 = -\left(1+{2mr\over\rho^2}\right) + 2 T_r \; {2mr\over\rho^2} + {\Delta\over \rho^2} \; T_r^2 + 
{T_\theta^2\over \rho^2}.
\end{equation}
This simplifies to
\begin{equation}
0 = -2mr + 4mr \;T_r + \Delta \; T_r^2 + T_\theta^2.
\end{equation}
But this has the obvious solution
\begin{equation}
T_\theta=0; \qquad  T_r = {-2mr \pm\sqrt{(2mr)^2 + (2mr)\Delta}\over\Delta}= 
 {-2mr \pm\sqrt{2mr(r^2+a^2)}\over\Delta}.
\end{equation}
Ultimately the sign $\pm$ of the square root will be chosen to distinguish a black hole from a white hole. 
Note that
\begin{eqnarray}
{-2mr \pm\sqrt{2mr(r^2+a^2)}\over\Delta} 
&=&
{-2mr \pm\sqrt{2mr(r^2+a^2)}\over\Delta}  \; 
{{-2mr \mp\sqrt{2mr(r^2+a^2)}} \over {-2mr \mp\sqrt{2mr(r^2+a^2)}} } 
\nonumber\\
&=& 
{-2mr\over  {-2mr \mp\sqrt{2mr(r^2+a^2)}} } 
\nonumber\\
&=& 
{2mr/(r^2+a^2)\over  {2mr/(r^2+a^2) \pm\sqrt{2mr/(r^2+a^2)}}}
\nonumber\\
&=& 
{\sqrt{2mr/(r^2+a^2)}\over  {\sqrt{2mr/(r^2+a^2)} \pm 1}} 
\nonumber\\
&=& 
\pm\;  {\sqrt{2mr/(r^2+a^2)}\over 1\pm  {\sqrt{2mr/(r^2+a^2)} }}.
\end{eqnarray}
That is, the relevant coordinate transformation can be recast as  
\begin{equation}
T_\theta=0; \qquad  T_r = \pm\; {\sqrt{2mr/(r^2+a^2)}\over 1 \pm\sqrt{2mr/(r^2+a^2)}}.
\end{equation}
So $T(r,\theta)$ is actually independent of $\theta$, and we explicitly have
\begin{equation}
T(r) = \pm \int{\sqrt{2mr/(r^2+a^2)}\over 1 \pm\sqrt{2mr/(r^2+a^2)}}
 \; dr.
\end{equation}
Thence
\begin{equation}
\bar t = t + T(r); \qquad d\bar t = dt + T_r; \qquad dt = d\bar t - T_r.
\end{equation}
That is, now suppressing the overbar, taking the Eddington--Finkelstein $t$-$r$  form of the Kerr metric and simply replacing
\begin{equation}
dt \to dt \mp  {\sqrt{2mr/(r^2+a^2)}\over 1 \pm\sqrt{2mr/(r^2+a^2)}}
 \; dr,
\end{equation}
will put the metric into unit-lapse form. Let us call the resulting line element the Eddington--Finkelstein-rain metric (to be abbreviated as EF-rain).
Explicitly
\begin{eqnarray}
\label{E:K3}
(\d s^2)_\EFrain &=& 
-\left( dt \mp  {\sqrt{2mr/(r^2+a^2)}\over 1 \pm\sqrt{2mr/(r^2+a^2)}}
 \; dr\right)^2
\\
&&
 + \d r^2 - 2 a \sin^2\theta \;\d r \; \d \phi + \rho^2\; \d\theta^2 
+ (r^2+a^2) \sin^2\theta \;\d\phi^2
\nonumber
\\
&&
+{2mr\over \rho^2}\; \left(\d t + \left[1  \mp  {\sqrt{2mr/(r^2+a^2)}\over 1 \pm\sqrt{2mr/(r^2+a^2)}}\right]
 \; dr  - a \sin^2\theta \; \d \phi\right)^2.\qquad
 \nonumber
\end{eqnarray}
Thence, this slightly simplifies to
\begin{eqnarray}
\label{E:K3}
(\d s^2)_\EFrain &=& 
-\left( dt \mp  {\sqrt{2mr/(r^2+a^2)}\over 1 \pm\sqrt{2mr/(r^2+a^2)}}
 \; dr\right)^2
\\
&&
 + \d r^2 - 2 a \sin^2\theta \;\d r \; \d \phi + \rho^2\; \d\theta^2 
+ (r^2+a^2) \sin^2\theta \;\d\phi^2
\nonumber
\\
&&
+{2mr\over \rho^2}\; \left(\d t +   {dr\over 1 \pm\sqrt{2mr/(r^2+a^2)}}
 - a \sin^2\theta \; \d \phi\right)^2.\qquad
 \nonumber
\end{eqnarray}
\enlargethispage{40pt}

Retrospectively checking that it is the upper sign that corresponds to a black hole,\footnote{This is most easily checked by setting $a\to0$ and comparing with the (black hole) Painlev\'e--Gullstrand form of the Schwarzschild line element.}  we have 
\begin{eqnarray}
\label{E:K3}
(\d s^2)_\EFrain &=& 
-\left( dt -  {\sqrt{2mr/(r^2+a^2)}\over 1 +\sqrt{2mr/(r^2+a^2)}}
 \; dr\right)^2
\\
&&
 + \d r^2 - 2 a \sin^2\theta \;\d r \; \d \phi + \rho^2\; \d\theta^2 
+ (r^2+a^2) \sin^2\theta \;\d\phi^2
\nonumber
\\
&&
+{2mr\over \rho^2}\; \left(\d t +   {dr\over 1 +\sqrt{2mr/(r^2+a^2)}}
  - a \sin^2\theta \; \d \phi\right)^2.\qquad
 \nonumber
\end{eqnarray}

In these Eddington--Finkelstein-rain coordinates the covariant metric 
is given by
\begin{equation}
(g_{ab})_\EFrain =
 {\left[\begin{array}{c|cc|c}
- 1+{2mr\over \rho^2} & g_{tr} &
 0 &- {2mar\over\rho^2}\sin^2\theta \\ \hline
 g_{tr} &g_{rr} &0&g_{r\phi}\\ 
 0&0&\;\;\rho^2\;\;&0\\ \hline
 -{2mar\over\rho^2}\sin^2\theta&g_{r\phi}&0&  \Sigma\,\sin^2\theta
\end{array}\right]}_{ab}
\end{equation}
subject to the relatively messy results that
\begin{equation}
g_{rr} = 1+  {a^2\sin^2\theta(2mr/\rho^2) \over
(r^2+a^2)(1+ \sqrt{2mr/(r^2+a^2)})^2};
\end{equation}
\begin{equation}
g_{tr} ={ {2mr/\rho^2} + \sqrt{2mr/(r^2+a^2)} \over 1+ \sqrt{2mr/(r^2+a^2)}};
\end{equation}
\begin{equation}
g_{r\phi} = -a \sin^2\theta \left( 1 +{2mr/\rho^2} + \sqrt{2mr/(r^2+a^2)}
\over 1 + \sqrt{2mr/( r^2+a^2)}\right).
\end{equation}

Remarkably, the inverse metric is again \emph{much} simpler
\begin{equation}
(g^{ab})_\EFrain =  {\left[\begin{array}{c|cc|c}
- 1&{\sqrt{2mr(r^2+a^2)}\over \rho^2}&
 0 & {\sqrt{2mra^2/(r^2+a^2)}\over\rho^2 (1+\sqrt{2mr/(r^2+a^2)})}\\ \hline
  {\sqrt{2mr(r^2+a^2)}\over \rho^2}&
 {\Delta\over\rho^2}
 &0& {a\over \rho^2}\\
 0&0&{1\over \rho^2}&0\\ \hline
 {\sqrt{2mra^2/(r^2+a^2)}\over\rho^2 (1+\sqrt{2mr/(r^2+a^2)})}& {a\over \rho^2}&0& {1\over \rho^2\sin^2\theta}
\end{array}\right]}^{ab}.
\end{equation}
\enlargethispage{30pt}
So we have again simplified the lapse, but again at the cost of complicating the flow vector, now in a slightly different manner:
\begin{equation}
N=1; \qquad (v^i)_\EFrain =-\left( {\sqrt{2mr(r^2+a^2)}\over \rho^2} ,\;
 0 ,\; {\sqrt{2mra^2/(r^2+a^2)}\over\rho^2 (1+\sqrt{2mr/(r^2+a^2)})}\right).
\end{equation}
Note that for the rain geodesics we again have $d\theta/dt=0$, so that $\theta(t)=\theta_\infty$ is again conserved. Furthermore we now have
\begin{equation}
\left({d\phi\over dr}\right)_\EFrain = {d\phi/dt\over dr/dt} ={ a\over(r^2+a^2)(1+\sqrt{2mr/(r^2+a^2)})}. 
\end{equation}
Therefore for these EF-rain geodesics we now have the relatively simple azimuthal behaviour
\begin{equation}
\phi(r) = \phi_\infty - \int_r^\infty { a\over(r^2+a^2)(1+\sqrt{2mr/(r^2+a^2)})}\; dr.
\end{equation}
Overall this EF-rain version of the Kerr spacetime is again quite straightforward, both  in terms of tractability and clarity of physical insight.

\subsection{Summary at this stage}

Up to this point, working only with $t$-coordinate transformations, we have  already constructed two novel and fully explicit unit-lapse versions of the Kerr spacetime, namely the BL-rain and EF-rain metrics. While establishing the existence of these BL-rain and EF-rain metrics is relatively easy, and the behaviour of the rain geodesics is transparent, these metrics can perhaps be further improved by working with $\phi$-coordinate transformations.

\section{Adjusting the flow vector}
\def\Doran{{\hbox{\tiny Doran}}}
\def\Natario{{\hbox{\tiny Nat\'ario}}}
\def\general{{\hbox{\tiny general}}}

Having now used the freedom in choosing the time coordinate to exhibit two explicit unit lapse forms of the Kerr solution, we shall consider the effects of using the freedom in choosing the azimuthal coordinate $\phi$ to further simplify the metric.  Remember that on quite general grounds we had seen that it is possible to transform the flow vector as follows
$v^\phi \to \bar v^\phi = v^\phi +\Phi_r v^r + \Phi_\theta v^\theta$. 
\begin{itemize}
\item 
In both of the specific examples we have investigated above, (BL-rain and EF-rain), one has $v^\theta=0$, so one might as well consider $v^\phi \to \bar v^\phi = v^\phi +\Phi_r v^r$. 
\item
In both of the specific examples we have investigated above, (BL-rain and\break  EF-rain), the only angular dependence in both the $v^r$ and $v^\phi$ components arises from a common factor of $\rho^{-2}$.
\item 
This suggests that it should be possible to eliminate $v^\phi$ completely by suitably choosing a coordinate transformation $\bar \phi = \phi + \Phi(r)$. 
\end{itemize}
We will now use this freedom to extract the Doran~\cite{doran} version of the Kerr spacetime metric via three distinct routes, from the BL-rain metric, from the EF-rain metric, and directly from the EF-null metric.
We shall also discuss Nat\'ario's version of the Kerr spacetime~\cite{natario}, wherein he does not set $v^\phi\to0$ but instead forces $v^\phi$ to be a very specific  function of $r$ and $\rho$.

\subsection{Doran metric: Route 1 (Boyer--Lindquist-rain)}

Let us start from the BL-rain (inverse) metric as explored above, 
\begin{equation}
(g^{ab})\BLrain =  {\left[\begin{array}{c|cc|c}
- 1 & {\sqrt{2mr(r^2+a^2)}\over \rho^2}  &
 0 & -{2mar\over \rho^2\Delta}\\ \hline
 {\sqrt{2mr(r^2+a^2)}\over \rho^2} &{\Delta\over \rho^2}
 &0&0\\
 0&0&{1\over \rho^2}&0\\ \hline
- {2mar\over \rho^2\Delta}&0&
0& 
{1-2mr/\rho^2\over\Delta\sin^2\theta}
\end{array}\right]}^{ab}.
\end{equation}

Recall that in these coordinates the flow vector is
\begin{equation}
(v^i)_\BLrain = \left( -{\sqrt{2mr(r^2+a^2)}\over \rho^2} ,\;
 0 ,\; {2mar\over \rho^2\Delta}\right).
\end{equation}
Now choose
\begin{equation}
\Phi_r = -\left(v^\phi\over v^r\right)_\BLrain  = {a \sqrt{2mr}  \over \Delta \sqrt{r^2+a^2}};
\qquad
\Phi(r) = \int {a \sqrt{2mr}  \over \Delta \sqrt{r^2+a^2}}\; dr.
\end{equation}
Then $\bar v^\phi\to 0$. However, in view of equation (\ref{E:inverse-metric-spatial}), and the fully explicit forms (\ref{E:inverse-metric-spatial-1})--(\ref{E:inverse-metric-spatial-3}),  the spatial part of the inverse 3-metric becomes slightly more complicated and we obtain (via this nonstandard route starting from the Boyer--Lindquist version of Kerr) the Doran~\cite{doran} form of the (inverse) Kerr metric
\begin{equation}
(g^{ab})_\Doran = \left[\begin{array}{c|cc|c}
-1 & {\sqrt{2mr(a^2+r^2)}\over \rho^2}  & 0 & 0\\
\hline
{\sqrt{2mr(a^2+r^2)}\over \rho^2} & {\Delta\over \rho^2}& 
0 & {a\sqrt{2mr\over a^2+r^2} \over \rho^2}\\
0 &0 & {1\over \rho^2} & 0\\
\hline
0 &  {a\sqrt{2mr\over a^2+r^2} \over \rho^2} & 
0 & {1\over (a^2+r^2) \sin^2\theta}
\end{array}\right]^{ab}.
\end{equation}
This is completely equivalent to starting with the Boyer--Lindquist form of Kerr and making the two coordinate transformations
\begin{equation}
dt \to dt -  {\sqrt{2mr(r^2+a^2)}\over\Delta}\; dr,
\end{equation}
\begin{equation}
d\phi\to d\phi-{a \sqrt{2mr}  \over \Delta \sqrt{r^2+a^2}} \;dr.
\end{equation}
Doing so results in
\enlargethispage{40pt}
\begin{eqnarray}
(d s^2)_\Doran &=& -d t^2 + \rho^2 \; d\theta^2 + (r^2+a^2)\sin^2\theta\;d\phi^2
\\
&& 
+
\left\{ 
{\rho d r\over\sqrt{r^2+a^2}}  + {\sqrt{2mr}\over \rho}\; (d t - a\sin^2\theta\; d \phi) 
\right\}^2.
\nonumber
\end{eqnarray}
The covariant metric is then\footnote{An easy consistency check is to set $a\to0$ and verify that one recovers the (black hole) Painlev\'e--Gullstrand version of the Schwarzschild spacetime.}
\begin{equation}
(g_{ab})_{Doran} = \left[\begin{array}{c|cc|c}
-1 +{2mr\over \rho^2} & \sqrt{2mr\over a^2+r^2} & 
0 & -{2mar\sin^2\theta\over \rho^2}\\
\hline
\sqrt{2mr\over a^2+r^2} & {\rho^2\over r^2+a^2}& 0 & -a\sqrt{2mr\over a^2+r^2} \sin^2\theta\\
0 &0 & \rho^2& 0\\
\hline
-{2mar\sin^2\theta\over \rho^2} & -a\sqrt{2mr\over a^2+r^2} \sin^2\theta & 
0 & \Sigma\, \sin^2\theta
\end{array}\right]_{ab}.
\end{equation}

\subsection{Doran metric: Route  2 (Eddington--Finkelstein-rain) }

Let us now start from the EF-rain (inverse) metric as explored above, 
\begin{equation}
(g^{ab})_\EFrain =  {\left[\begin{array}{c|cc|c}
- 1&{\sqrt{2mr(r^2+a^2)}\over \rho^2}&
 0 & {\sqrt{2mra^2/(r^2+a^2)}\over\rho^2 (1+\sqrt{2mr/(r^2+a^2)})}\\ \hline
  {\sqrt{2mr(r^2+a^2)}\over \rho^2}& {\Delta\over\rho^2}
 &0& {a\over \rho^2}\\
 0&0&{1\over \rho^2}&0\\ \hline
 {\sqrt{2mra^2/(r^2+a^2)}\over\rho^2 (1+\sqrt{2mr/(r^2+a^2)})}& {a\over \rho^2}&0& {1\over \rho^2\sin^2\theta}
\end{array}\right]}^{ab}.
\end{equation}
In these coordinates the flow vector is 
\begin{equation}
(v^i)_\EFrain = -\left( {\sqrt{2mr(r^2+a^2)}\over \rho^2} ,\;
 0 ,\; {\sqrt{2mra^2/(r^2+a^2)}\over\rho^2 (1+\sqrt{2mr/(r^2+a^2)})}\right).
\end{equation}
Now choose   
\begin{equation}
\Phi_r = -\left(v^\phi\over v^r\right)_{EF-rain}  
=- {a /(r^2+a^2)  \over 1+ \sqrt{2mr/(r^2+a^2)}}.
\end{equation}
So that
\begin{equation}
\Phi(r) = -\int {a /(r^2+a^2) \over 1+\sqrt{2mr/(r^2+a^2)}}\; dr.
\end{equation}

Then $\bar v^\phi\to 0$. However, in view of equation (\ref{E:inverse-metric-spatial}), and the fully explicit forms (\ref{E:inverse-metric-spatial-1})--(\ref{E:inverse-metric-spatial-3}),  the spatial part of the inverse 3-metric becomes slightly more complicated and we again obtain the Doran form of the (inverse) Kerr metric
\begin{equation}
(g^{ab})_\Doran = \left[\begin{array}{c|cc|c}
-1 & {\sqrt{2mr(a^2+r^2)}\over \rho^2}  & 0 & 0\\
\hline
{\sqrt{2mr(a^2+r^2)}\over \rho^2} & {\Delta\over \rho^2}& 
0 & {a\sqrt{2mr\over a^2+r^2} \over \rho^2}\\
0 &0 & {1\over \rho^2} & 0\\
\hline
0 &  {a\sqrt{2mr\over a^2+r^2} \over \rho^2} & 
0 & {1\over (a^2+r^2) \sin^2\theta}
\end{array}\right]^{ab}.
\end{equation}
This is completely equivalent to starting with the Eddington--Finkelstein $t$-$r$ form of Kerr and making the two coordinate transformations
\begin{equation}
dt \to dt -  {\sqrt{2mr/(r^2+a^2)}\over 1 +\sqrt{2mr/(r^2+a^2)}}
 \; dr,
\end{equation}
\begin{equation}
d\phi\to d\phi-{a /(r^2+a^2) \over 1+ \sqrt{2mr/(r^2+a^2)}} \;dr.
\end{equation}
Doing so again results in~\cite{doran}
\begin{eqnarray}
(d s^2)_\Doran &=& -d t^2 + \rho^2 \; d\theta^2 + (r^2+a^2)\sin^2\theta\;d\phi^2
\\
&& 
+
\left\{ {\rho\,d r\over\sqrt{r^2+a^2}} + {\sqrt{2mr}\over \rho}\; (d t - a\sin^2\theta\; d \phi) \right\}^2.
\nonumber
\end{eqnarray}

\subsection{Doran metric: Route  3 (Eddington--Finkelstein-null)}
\enlargethispage{30pt}
The original way of getting to the Doran metric~\cite{doran} was to take 
the  ``advanced Eddington--Finkelstein null coordinate'' version of the Kerr solution~\cite{Kerr},
(with $a\to-a$ to conform with standard conventions):
\begin{eqnarray}
(d s^2)_\EFnull &=& -\left[ 1 - {2mr\over r^2+a^2\cos^2\theta}\right]\; \left(d u - a \sin^2\theta \; d \phi\right)^2
\nonumber
\\
&&
+2 \left(d u - a \sin^2\theta \; d \phi\right) \; \left(d r - a \sin^2\theta \; d \phi\right)
\nonumber
\\
&&
+ (r^2+a^2\cos^2\theta)\; (d\theta^2+\sin^2\theta\;d\phi^2).
\label{E:K1}
\end{eqnarray}
One then simultaneously makes the two $m$-dependent coordinate transformations~\cite{doran}
\begin{equation}
d u = d t + {d r\over 1+ \sqrt{2mr/(r^2+a^2)}}; 
\end{equation}
\begin{equation}
d\phi_\mathrm{Doran} = d\phi + {a\;d r\over r^2+a^2+\sqrt{2mr(r^2+a^2)}}.
\end{equation}
This is of course equivalent to first applying the $u$ transformation to go to from EF-null to EF-rain coordinates, and then subsequently applying the $\phi$ transformation to go from EF-rain coordinates to Doran coordinates.
After dropping the subscript ``Doran'', in the new $(t,r,\theta,\phi)$ coordinates Doran's version of the Kerr line element again takes the form:
\begin{eqnarray}
(d s^2)_\Doran &=& -d t^2 + \rho^2 \; d\theta^2 + (r^2+a^2)\sin^2\theta\;d\phi^2
\\
&& 
+
\left\{ {\rho\,d r\over\sqrt{r^2+a^2}} + {\sqrt{2mr}\over \rho}\; (d t - a\sin^2\theta\; d \phi) \right\}^2.
\nonumber
\end{eqnarray}
From the line element it is easy to extract $g_{ab}$ the matrix of metric components. 
Explicitly
\begin{equation}
(g_{ab})_\Doran = \left[\begin{array}{c|cc|c}
-1 +{2mr\over \rho^2} & \sqrt{2mr\over a^2+r^2} & 
0 & -{2mar\sin^2\theta\over \rho^2}\\
\hline
\sqrt{2mr\over a^2+r^2} & {\rho^2\over r^2+a^2}& 0 & -a\sqrt{2mr\over a^2+r^2} \sin^2\theta\\
0 &0 & \rho^2& 0\\
\hline
-{2mar\sin^2\theta\over \rho^2} & -a\sqrt{2mr\over a^2+r^2} \sin^2\theta & 
0 & \Sigma\, \sin^2\theta
\end{array}\right]_{ab}.
\end{equation}

It is easy to extract invert $g_{ab}$ to obtain $g^{ab}$ the matrix of inverse-metric components. 
Explicitly
\begin{equation}
(g^{ab})_\Doran = \left[\begin{array}{c|cc|c}
-1 & {\sqrt{2mr(a^2+r^2)}\over \rho^2}  & 0 & 0\\
\hline
{\sqrt{2mr(a^2+r^2)}\over \rho^2} & {\Delta\over \rho^2}& 
0 & {a\sqrt{2mr\over a^2+r^2} \over \rho^2}\\
0 &0 & {1\over \rho^2} & 0\\
\hline
0 &  {a\sqrt{2mr\over a^2+r^2} \over \rho^2} & 
0 & {1\over (a^2+r^2) \sin^2\theta}
\end{array}\right]^{ab}.
\end{equation}
Note in particular that $g^{tt}=-1$ as claimed. Note  that the shift vector
\enlargethispage{20pt}
\begin{equation}
(v^i)_\Doran = - \left(  {\sqrt{2mr(a^2+r^2)}\over \rho^2} ,0,0\right)
\end{equation}
is particularly simple. Finally with symbolic manipulation software it is easy to check that the metric is indeed Ricci flat $R_{ab}=0$. 

Of the three distinct routes for getting to the Doran metric~\cite{doran}, the EF-null route is traditional, 
but the BL-rain and EF-rain routes are perhaps more informative, and provide us with additional insight.
Overall, we feel that the BL-rain route (BL $\to$ BL-rain $\to$ Doran) is in many ways the simplest route --- of course one has to get to the BL metric in the first place. 

\subsection{Rain geodesics in the Doran metric}

However one gets to the Doran metric, the rain geodesics are just integral curves of the flow vector field
\begin{equation}
(v^i)_\Doran = - \left(  {\sqrt{2mr(a^2+r^2)}\over \rho^2} ,0,0\right).
\end{equation}
But this now implies that both $\theta$ and $\phi$ are constant along the Doran rain geodesics --- effectively one has simplified the azimuthal evolution of the rain geodesics by craftily picking an azimuthal coordinate transformation to strategically cancel the azimuthal evolution occurring in the rain geodesics as expressed in either BL-rain or EF-rain coordinates. 

In these Doran coordinates the rain geodesics satisfy
\begin{equation}
t(\tau) = \tau; \qquad \theta(\tau)= \theta_\infty; \qquad \phi(\tau) = \phi_\infty;
\end{equation}
while
\begin{equation}
{d r\over dt} =  -{\sqrt{2mr(a^2+r^2)}\over r^2+a^2\cos^2\theta_\infty}.
\end{equation}
So formally at least
\begin{equation}
t = t_0 -
\int_{r_0}^r {r^2+a^2\cos^2\theta_\infty\over \sqrt{2mr(a^2+r^2)}} \;\; dr.
\end{equation}
Unfortunately, performing this integral involves an incomplete Elliptic integral of the first kind, so the function $t(r)$ and its inverse $r(t)$ are at best implicit rather than fully explicit.

\subsection{Nat\'ario version of the Kerr spacetime}

Yet another unit-lapse version of the Kerr spacetime has been provided by Nat\'ario in reference~\cite{natario}:
\begin{equation}
(ds^2)_\Natario = - dt^2 + {\rho^2\over\Sigma} (dr-v \, dt)^2 + \rho^2 d\theta^2 
+ \Sigma \sin^2\theta\left( d\phi + \delta d\theta-\Omega dt\right)^2.
\end{equation}
Natario started from Boyer--Lindquist coordinates and then invoked the further coordinate transformations
\begin{equation}
d\bar t = dt -  {\sqrt{2mr(r^2+a^2)}\over\Delta}\; dr,
\end{equation}
\begin{equation}
d\bar \phi = d\phi + \Phi_r \; dr + \Phi_\theta \;d\theta.
\end{equation}
Now the $t$ coordinate transformation, considered by itself, simply brings the BL metric into the BL-rain form previously considered. But the $\phi$ transformation Nat\'ario used did \emph{not} then bring the metric into Doran form --- instead Nat\'ario chose to enforce 
\begin{equation}
(v^\phi)_\Natario = \Omega = {2mra\over\rho^2\Sigma},
\end{equation}
where as previously
\begin{equation}
\Sigma = r^2+a ^2 +{2mra^2\over\rho^2} \sin^2\theta
=\rho^2 + a^2 \left(1+{2mr\over\rho^2} \right)\sin^2\theta.
\end{equation}
Natario's choice for $v^\phi$ leads to a rather complicated expression for $\Phi(r,\theta)$. \\
%
Specifically, starting from
\begin{equation}
\Phi_r = {(v^\phi)_\Natario - (v^\phi)_\BLrain \over (v^r)_\BLrain}=
 {(v^\phi)_\Natario - (v^\phi)_\BLrain \over (v^r)_\Natario},
\end{equation}
and then substituting and integrating, one can formally extract $\Phi(r,\theta)$ ---  but the result is not particularly edifying.
\enlargethispage{30pt}
In contrast 
\begin{equation}
v = - {\sqrt{2mr(r^2+a^2)}\over \rho^2},
\end{equation}
is quite tractable.

Unfortunately the quantity $\delta(r,\theta)$ is quite intractable:
\begin{equation}
\delta(r,\theta) = -a^2 \sin(2\theta) \int_r^\infty {v\Omega\over\Sigma} dr.
\end{equation}
Explicitly
\begin{equation}
\delta(r,\theta) = -a^2 \sin(2\theta) \int_r^\infty {
2mar \sqrt{2mr(r^2+a^2)} \over [ (r^2+a^2) (r^2+a^2 \cos^2\theta) + 2 \sin^2\theta ma^2 r ]^2
} \;dr.
\end{equation}
The integration leads to incomplete Elliptic integrals, so the presence of $\delta(r,\theta)$ \emph{in the line element}  implies the implicit presence of incomplete Elliptic integrals \emph{in the metric components} themselves. This renders Nat\'ario's form of the metric for the Kerr spacetime less attractive than it first appears.
\enlargethispage{40pt}

For completeness we point out that
\begin{equation}
(g_{ab})_\Natario = \left[\begin{array}{c|cc|c}
-1 +{\rho^2 v^2\over\Sigma} + \Sigma \sin^2\theta \Omega^2 &
-{\rho^2 v\over\Sigma} & 
- \delta \Sigma \sin^2\theta\Omega
&-\Sigma \sin^2\theta\Omega\\ \hline
-{\rho^2 v\over\Sigma} & {\rho^2\over\Sigma} & 0 &0\\
 - \delta \Sigma \sin^2\theta\Omega &0 
 & \rho^2 + \delta^2 \Sigma \sin^2\theta 
 &  \delta \Sigma \sin^2\theta\\ \hline
 - \Sigma \sin^2\theta\Omega &0 & \delta \Sigma \sin^2\theta &
 \Sigma\sin^2\theta
\end{array}\right]_{ab}.
\end{equation}

The metric determinant is again
\begin{equation}
\det\left( (g_{ab})_\Natario \right) = - \rho^4 \sin^2\theta,
\end{equation}
as it should be. 
(The relevant Jacobi matrices are all determinant unity.) 

Finally the inverse metric is 
\begin{equation}
(g^{ab})_\Natario = \left[\begin{array}{c|cc|c}
-1  & -v & 0 & -\Omega
\\ \hline
-v& {\Sigma\over \rho^2}- v^2 & 0 &-\Omega v\\
 0 &0 
 & {1\over\rho^2} 
 & - {\delta \over \rho^2}\\ \hline
 - \Omega &-\Omega v &-{ \delta\over\rho^2} &
{1\over \Sigma\sin^2\theta}+ {\delta^2\over\rho^2}  -\Omega^2 
\end{array}\right]^{ab}.
\end{equation}
As required, the lapse function is indeed unity and the flow vector is now
\begin{equation}
(v^i)_\Natario = (v,0,\Omega). 
\end{equation}

For rain geodesics in the Nat\'ario metric $\theta$ is again conserved, so that $\theta(r)=\theta_\infty$.\\
 In addition
\begin{equation}
\left(d\phi\over dr\right)_\Natario = {d\phi/dt\over dr/dt} = {\Omega\over v} = 
-  \sqrt{2mr\over r^2+a^2} \; {a\over \Sigma}.
\end{equation}
This leads to the intractable integral
\begin{equation}
\phi(r) = \phi_\infty + \int_r^\infty  \sqrt{2mr\over r^2+a^2} \; {a\over r^2+a^2 + {2mra^2\over r^2+a^2\cos^2\theta} \sin^2\theta} \; dr.
\end{equation}
The only other significant drawback of the Nat\'ario form of the metric for Kerr spacetime  is the explicit presence of the quantity $\delta(r,\theta)$ \emph{in the metric components}, hiding the implicit presence of several incomplete Elliptic integrals. 

\subsection{Summary at this stage}

Up to this point, first working only with  $t$-coordinate transformations, we have constructed two novel and fully explicit unit-lapse versions of the Kerr spacetime, namely the BL-rain and EF-rain metrics. Then
with certain specific choices for the $\phi$-coordinate transformations
have recovered the fully explicit Doran~\cite{doran} and semi-explicit Nat\'ario~\cite{natario} metrics. 
While establishing the existence of all four of these unit-lapse metrics is relatively easy,  it does open the question of what the most general unit-lapse version of the Kerr spacetime might look like.

\section{General unit-lapse representation of the Kerr metric}

Given what we have seen so far, the development of a general unit-lapse representation of the Kerr metric is now straightforward --- pick \emph{any} one of the four specific unit-lapse metrics we have investigated (BL-rain, EF-rain, Doran, Nat\'ario) and 
for an \emph{arbitrary} function $\Phi(r,\theta)$ simply transform the $\phi$ coordinate $\phi\to\bar\phi-\Phi(r,\theta)$, while leaving the $t$ coordinate intact. 
That is, replace 
\begin{equation}
d\phi \to d\phi -\Phi_r \, dr - \Phi_\theta\, \d\theta
\end{equation}
in the line element.
Let us explicitly do this for the Doran line element.  We find
\begin{eqnarray}
(d s^2)_\general &=& -d t^2 + \rho^2 \; d\theta^2 + (r^2+a^2)\sin^2\theta\;(d\phi -\Phi_r \, dr - \Phi_\theta\, \d\theta)^2
\\
&& 
+
\left\{ {\rho d r\over\sqrt{r^2+a^2}}  + {\sqrt{2mr}\over \rho}\; (d t - a\sin^2\theta\; (d\phi -\Phi_r \, dr - \Phi_\theta\, \d\theta)) \right\}^2.
\nonumber
\end{eqnarray}
Let us write
\begin{equation}
(g_{ab})_\general = (g_{ab})_\Doran  + \Delta_1 (g_{ab}) + \Delta_2 (g_{ab}).
\end{equation}
We have already calculated $(g_{ab})_\Doran$. 

\clearpage
The first-order and second-order shifts, (linear and quadratic in the gradients of $\Phi$), are:
\begin{equation}
\Delta_1 (g_{ab}) = \sin^2\theta \left[\begin{array}{c|cc|c}
0 & {2mar\over\rho^2}  \Phi_r& {2mar\over\rho^2}  \Phi_\theta
 & 0\\
\hline
{2mar\over\rho^2}  \Phi_r &
 2a\sqrt{2mr\over r^2+a^2}  \Phi_r & 
 a\sqrt{2mr\over r^2+a^2} \Phi_\theta & -\Sigma \Phi_r\\
{2mar\over\rho^2}  \Phi_\theta & a\sqrt{2mr\over r^2+a^2} \Phi_\theta  
& 0&  -\Sigma \Phi_\theta\\
\hline
0&  -\Sigma \Phi_r &  -\Sigma \Phi_\theta &
0
\end{array}\right]_{ab}.
\end{equation}
and 
\begin{equation}
\Delta_2 (g_{ab}) = \Sigma \sin^2 \theta \left[\begin{array}{c|cc|c}
0 &0&0& 0\\
\hline
0&
 \Phi_r^2 & 
 \Phi_r \Phi_\theta &0\\
0&\Phi_r \Phi_\theta&\Phi_\theta^2&  0\\
\hline
0&  0 &  0& 0
\end{array}\right]_{ab} = \Sigma\sin^2\theta \; \Phi_a \Phi_b.
\end{equation}
Note that only \emph{some} of the components of $(g_{ab})_\Doran$ change, 
and that they do so in a quite well-controlled manner.

It is straightforward to now invert $(g_{ab})_\general$ to obtain $(g^{ab})_\general$ the matrix of inverse-metric components. 
Let us write
\begin{equation}
(g^{ab})_\general = (g^{ab})_\Doran  + \Delta_1 (g^{ab}) + \Delta_2 (g_{ab}).
\end{equation}
We have already calculated $(g^{ab})_\Doran$. 

The first-order and second-order shifts are:
\begin{equation}
\Delta_1 (g^{ab}) = {1\over\rho^2} \left[\begin{array}{c|cc|c}
0 &0&0&  {\sqrt{2mr(r^2+a^2)}} \; \Phi_r \\
\hline
0 & 0 &0  & \Delta\; \Phi_r\\
0 & 0 & 0& {\Phi_\theta} \\
\hline
{\sqrt{2mr(r^2+a^2)}} \;\Phi_r&  \Delta\; \Phi_r  &\;\; {\Phi_\theta} \;\;&
2a\sqrt{2mr\over r^2+a^2} \Phi_r
\end{array}\right]_{ab}.
\end{equation}
and 
\begin{equation}
\Delta_2 (g^{ab}) = {\Delta \; \Phi_r^2 + \Phi_\theta^2\over \rho^2} \left[\begin{array}{c|cc|c}
0 &0&0& 0\\
\hline
0&
0 & 
0 &0\\
0&0&0&  0\\
\hline
0&  0 &  0& 1
\end{array}\right]^{ab}.
\end{equation}
Note that only \emph{some} of the components of $(g^{ab})_\Doran$ change, 
and that they do so in a well-controlled manner.
\enlargethispage{20pt}

This represents the most general unit-lapse representation of the Kerr spacetime geometry, 
keeping the $(r,\theta)$ coordinates in the usual spherical oblate spheroidal form.
Note that, as advertised, $\phi$-coordinate transformations that manifestly preserve the stationary axisymmetric nature of the spacetime, while also preserving the $(r,\theta)$ spherical oblate spheroidal coordinates, can be used to adjust the flow vector at the price of also affecting the 3-metric.

Adding $(r,\theta)$ coordinate transformations to the discussion does not seem to add much to the physics --- the  $(r,\theta)$ spherical oblate spheroidal coordinates seem to be preferred coordinates --- though this seems to be more than just an effect of stationarity and axisymmetry. 
There seems to be more at play here, and we hope to address these issues in future work.


\section{Conclusions}

What have we learned from this discussion?
First, unit lapse versions of stationary spacetimes are extremely useful in that they immediately provide a class of timelike geodesics, the ``rain geodesics'' (zero angular momentum observers, ZAMOs, that are dropped from spatial infinity with zero initial velocity), that provide an explicit  and tractable probe of the spacetime physics. 
Second, the Kerr spacetime (which is an exact solution of the vacuum Einstein equations that is the default option for describing astrophysically interesting black holes) admits an infinite class of unit-lapse coordinate charts. 
The Doran coordinates are one example, but so are the Nat\'ario coordinates, 
as are the BL-rain and EF-rain coordinates introduced herein. 

Improved coordinate systems for the Kerr spacetime are strategically and tactically important for a better understanding of the technically challenging and astrophysically important Kerr spacetime. 
See for instance attempts at finding a ``Gordon form'' for the Kerr spacetime~\cite{Gordon-form}, 
and attempts at upgrading the ``Newman--Janis trick'' from an ansatz to an algorithm~\cite{ansatz}.
Finally we should also mention that the discussion herein also impacts the observational ability to distinguish exact Kerr black holes from various ``black hole mimickers'' --- see for instance references~\cite{small-dark-heavy, BH-in-GR}, and more recently references~\cite{phenomenology,viability,geodesicaly-complete,pandora, causal,LISA}, and references~\cite{Simpson:2020,Simpson:2019,Simpson:2018,Boonserm:2018,Simpson:2019-core, Bardeen:1968,Hayward:2005,Frolov:2014,Frolov:2017}.

\section*{Acknowledgements}

JB was supported by a MSc scholarship funded by the Marsden Fund, 
via a grant administered by the Royal Society of New Zealand.
\\
TB was supported by a Victoria University of Wellington MSc scholarship, 
and was also indirectly supported by the Marsden Fund, 
via a grant administered by the Royal Society of New Zealand.
\\
AS was supported by a Victoria University of Wellington PhD Doctoral Scholarship,
and was also indirectly supported by the Marsden fund, 
via a grant administered by the Royal Society of New Zealand.
\\
MV was directly supported by the Marsden Fund, 
via a grant administered by the Royal Society of New Zealand.


\end{document}